\input amstex
\magnification=\magstep1 
\baselineskip=13pt
\documentstyle{amsppt}
\vsize=8.7truein \CenteredTagsOnSplits \NoRunningHeads
\def\ii{\bold{i}}
\def\EE{\bold {E\thinspace}}
\def\dist{\operatorname{dist}}
\def\FF{\Cal{F}}
\topmatter
 
\title  On the dependence of the zero-free region of a partition function on the external field \endtitle 
\author Alexander Barvinok  \endauthor
\address Department of Mathematics, University of Michigan, Ann Arbor,
MI 48109-1043, USA \endaddress
\email barvinok$\@$umich.edu \endemail
\date August 12,   2026 \enddate
\thanks  \endthanks 
\abstract Let $\{0, 1\}^n$ be the Boolean cube, endowed with the probability product measure, where ${\Bbb P}(1)=p$ and ${\Bbb P}(0)=q$ with $0 < p \leq q$ and 
$p+q=1$. For $i=1, \ldots, m$, let
$\phi_i: \{0, 1\}^n \longrightarrow {\Bbb C}$ be $L_i$-Lipschitz functions in the Hamming metric, such that each $\phi_i$ depends on 
at most $r$ coordinates of $x \in \{0, 1\}^n$, where $rp \geq 12$. For $j=1, \ldots, n$, let $I_j $ be the set of indices $i$ such that $\phi_i$ depends on the $j$-th coordinate. We prove that $\EE \exp\left\{ \sum_{i=1}^m \phi_i \right\} \ne 0$ provided 
$\sum_{i \in I_j} L_i \leq {1 \over 10 \sqrt{rp}}$ for all $j$. This translates into a regime for $\pm 1$ spin systems, where a linear increase in the energy of multi-spin interactions requires only a logarithmic increase of the external field to keep the partition function zero-free and the system away from the phase transition. As a corollary, we obtain efficient deterministic algorithms to approximate the partition function in the zero-free region.
 \endabstract
\endtopmatter
\document

\head 1. Introduction and the main results \endhead 

We discuss how the presence of an external field affects the zero-free region of the partition function in a $\pm 1$ spin model. This is a classical topic that has been the subject of an extensive research, see \cite{FV18} for a review and references. Here we are interested in the situation of multi-spin interactions, that is when the energy of the system is the sum of the energies of subsystems involving some fixed number $r \geq 2$ of spins, and we are interested in the case of $r \gg 2$. We describe what appears to be a new type of the behavior for such a large $r$: there is a range for which a linear increase in the energy of interactions requires only a logarithmic increase of the external field to keep the partition function zero-free. Once the external field becomes strong enough, the system switches to a more familiar type of the behavior, where a linear increase of the energy of interactions requires a likewise linear increase of the external field, to maintain zero-freeness.

Below, we first state the main results in combinatorial and probabilistic terms, and then provide a statistical physics interpretation.

\subhead (1.1) The setup \endsubhead Let $\{0, 1\}^n$ be the Boolean cube of all $n$-vectors $x=\left(\xi_1, \ldots, \xi_n \right)$, where $\xi_j \in \{0, 1\}$ for 
$j=1, \ldots, n$. We consider the {\it Hamming metric} in $\{0, 1\}^n$, where the distance between any two vectors is the number of the coordinates where they differ:
$$\dist(x, y)=\left| j: \ \xi_j \ne \eta_j \right| \quad \text{where} \quad x=\left(\xi_1, \ldots, \xi_n\right) \quad \text{and} \quad y=\left(\eta_1, \ldots, \eta_n\right).$$
We consider complex-valued functions $\phi: \{0, 1\}^n \longrightarrow {\Bbb C}$. For $L >0$, we say that $\phi$ is $L$-{\it Lipschitz}, provided
$$|\phi(x)-\phi(y)| \ \leq \ L \dist(x, y) \quad \text{for all} \quad x, y \in \{0, 1\}^n.$$
For $J \subset \{1, \ldots, n\}$, we say that $\phi$ {\it depends on the coordinates} $\{\xi_j:\ j \in J\}$, provided 
$$\phi\left(\xi_1, \ldots, \xi_n\right) = \phi\left(\eta_1, \ldots, \eta_n \right) \quad \text{whenever} \quad \xi_j = \eta_j \quad \text{for all} \quad j \in J,$$
and $J$ is the minimal set under inclusion with this property.
 We say that $\phi$ {\it depends on at most} $r$ {\it coordinates}, provided 
 $|J| \leq r$.
 
 We choose numbers $0 < p \leq q$ such that $p+q=1$ and make $\{0, 1\}^n$ a probability space by letting 
 $${\Bbb P}(x)=\prod_{j=1}^n p^{\xi_j} q^{1-\xi_j} \quad \text{provided} \quad x=\left(\xi_1, \ldots, \xi_n\right). \tag1.1.1$$
 In other words, the coordinates $\xi_j$ of a point $x \in \{0, 1\}^n$ are independent random variables such that
 $${\Bbb P}(\xi_j=1)=p \quad \text{and} \quad {\Bbb P}(\xi_j=0)=q \quad \text{for} \quad j=1, \ldots, n.$$
 We consider expectations of the type 
 $$\EE \exp\left\{ \sum_{i=1}^m \phi_i \right\}, \tag1.1.2$$
 where $\phi_i: \{0, 1\}^n \longrightarrow {\Bbb C}$ are $L_i$-Lipschitz functions, each depending on at most $r$ coordinates. We are interested in the conditions that guarantee that (1.1.2) is not 0, in the regime where $m$ and $n$ are allowed to grow but $r, p$ and $q$ remain fixed.
 
 Now we are ready to state the first main result. 
 
 \proclaim{(1.2) Theorem} Let $\{0, 1\}^n$ be the Boolean cube endowed with the probability product measure 
$${\Bbb P}(x)=\prod_{j=1} p^{\xi_j} q^{1-\xi_j} \quad \text{for} \quad x=\left(\xi_1, \ldots, \xi_n\right),$$
where $0 < p \leq q$ and $p+q=1$. 

Let $\phi_1, \ldots, \phi_m: \{0, 1\}^n \longrightarrow {\Bbb C}$ be functions with respective Lipschitz constants $L_1, \ldots, L_m$ in the Hamming metric of 
$\{0, 1\}^n$.
For $i=1, \ldots, m$, suppose that $\phi_i$ depends on the coordinates 
$\left\{ \xi_j: \ j \in J_i \right\}$ for some $J_i \subset \{1, \ldots, n\}$. Suppose further that 
\roster
\item We have $|J_i| \ \leq \ r$ for some such that $rp \geq 12$ and all $i=1, \ldots, m$;
\item We have 
$$\sum_{i:\ j \in J_i} L_i \ \leq \ {1 \over 10 \sqrt{rp}} \quad \text{for} \quad j=1, \ldots, n;$$
\endroster
Then 
$$\EE \exp\left\{ \sum_{i=1}^m  \phi_i \right\} \ne 0.$$
\endproclaim
Thus we assume that every function $\phi_i$ depends on at most $r$ coordinates. The sum 
$$\sum_{i:\ j \in J_i} L_i$$
bounds the total influence of the $j$-th coordinate $\xi_j$ in (1.1.2). We note that as long as $r$ stays fixed while $p$ is decreasing, the bound for the influence that ensures the conclusion of Theorem 1.2
increases as $p^{-1/2}$, as long as $rp$ remains sufficiently large. For a fixed $p$, as $r$ increases, the bound decreases as $r^{-1/2}$, and this dependence on $r$ is optimal up to a constant \cite{Ba26}.

Once $rp$ gets small, we observe a different type of the behavior.

 \proclaim{(1.3) Theorem} Let $\{0, 1\}^n$ be the Boolean cube endowed with the probability product measure 
$${\Bbb P}(x)=\prod_{j=1} p^{\xi_j} q^{1-\xi_j} \quad \text{for} \quad x=\left(\xi_1, \ldots, \xi_n\right),$$
where $0 < p \leq q$ and $p+q=1$.

 Let $\phi_1, \ldots, \phi_m: \{0, 1\}^n \longrightarrow {\Bbb C}$ be functions with respective Lipschitz constants $L_1, \ldots, L_m$ in the Hamming metric of 
$\{0, 1\}^n$, and let $L >0$ be real number.

 For $i=1, \ldots, m$, suppose that $\phi_i$ depends on the coordinates 
$\left\{ \xi_j: \ j \in J_i \right\}$ for some $J_i \subset \{1, \ldots, n\}$. Suppose further that 
\roster
\item We have $|J_i| \ \leq \ r$ for some $r \geq 1$ and all $i=1, \ldots, m$;
\item We have 
$$\sum_{i:\ j \in J_i} L_i \ \leq \ {1 \over 5} L  \quad \text{for} \quad j=1, \ldots, n;$$
\item We have 
$$p={e^{-6L} \over r}.$$
\endroster
Then 
$$\EE \exp\left\{ \sum_{i=1}^m \phi_i \right\} \ne 0.$$
\endproclaim

After some preparations in Section 3, we prove Theorem 1.2 in Section 4 and Theorem 1.3 in Section 5.

\subhead (1.4) Statistical physics interpretation \endsubhead We have $n$ particles, and the $j$-th particle has $\xi_j=\pm 1$ {\it spin}. A an assignment 
$x=\left(\xi_1, \ldots, \xi_n\right)$ of spins is a {\it configuration} and the cube $\{-1, 1\}^n$ is interpreted as the set of all possible configurations on $n$ particles.
Spins interact in $m$ groups, and the contribution of the $i$-th group to the {\it energy} of the configuration is proportional to $\phi_i(x)$. A number $\alpha \in {\Bbb R}$ represents the {\it external field}, so that the contribution of the $j$-th spin to the energy of the configuration is represented by $-\alpha \xi_j$.

The function $H: \{-1, 1\}^n \longrightarrow {\Bbb R}$,
$$H(x)=-\sum_{i=1}^m \phi_i(x) + \alpha \sum_{j=1}^n\xi_j \quad \text{where} \quad x=\left(\xi_1, \ldots, \xi_n\right),$$
is called the {\it Hamiltonian} of the system. Compared to the standard definition, cf. \cite{FV18}, we made a sign change in $\alpha$. The change is made for convenience and is not substantive, as one can apply the transformation $\xi_j \longmapsto -\xi_j$ and redefine $\phi_i$ accordingly.

  There is a real parameter $\beta >0$, interpreted as the {\it inverse temperature}, so the total energy of the systems is 
$\beta H$. The sum
$$Z(H, \beta)=\sum_{x \in \{-1, 1\}^n } e^{-\beta H(x)} = \sum_{x \in \{-1, 1\}^n} \exp\left\{ \beta \sum_{i=1}^m \phi_i(x) - \beta \alpha \sum_{j=1}^n \xi_j \right\} \tag1.4.1$$ is 
the {\it partition function} of the system, and it makes $\{-1, 1\}^n$ a probability space with the {\it Gibbs measure}
$${\Bbb P}(x) = {e^{-\beta H(x)} \over Z(H, \beta)} \quad \text{for} \quad x \in \{-1, 1\}^n,$$
so that lower energy configurations are more likely, see, for example, \cite{FV18}. 

So far, all parameters in (1.4.1) are {\it real}. The idea to extend (1.4.1) to {\it complex} parameters belongs to Lee and Yang, who showed in \cite{YL52} that the 
complex roots of the function 
$$\alpha  \longmapsto Z(H, \beta), \quad \alpha \in {\Bbb C}, \tag1.4.2$$
in (1.4.1) are responsible for {\it phase transitions}, when one takes the {\it thermodynamic limit} as $n \longrightarrow \infty$, see \cite{FV18}. The idea is that as $n$ grows and the system grows in some regular way, complex roots of (1.4.2) may converge to real values, making the {\it pressure} 
$$\alpha \longmapsto {1 \over n \beta} \ln Z(H, \beta)$$
to lose analyticity in the limit. If the roots stay away from the real axis, the pressure remains analytic in the limit, and there is no phase transition.

In \cite{LY52}, Lee and Yang famously showed that there is no phase transition at positive temperatures in the {\it ferromagnetic Ising model} with non-zero constant external fields. The model corresponds to the situation where the interactions $\phi_i$ are indexed by unordered pairs $\{j_1, j_2\}$ of spins with $\phi_{j_1 j_2}(x)=b_{j_1 j_2} \xi_{j_1} \xi_{j_2}$ for some $b_{j_1 j_2} \geq 0$ (hence the model is called ferromagnetic, assigning lower energy to aligned spins and making configurations of aligned spins more likely), 
 and the Lee - Yang Theorem states that as long as $\beta >0$, the complex zeros of the univariate function
 $$\alpha \longmapsto Z(H, \beta)$$ lie on the line $\Re\thinspace \alpha=0$. The Lee - Yang Theorem was extended by Suzuki and Fisher \cite{SF71} to the case of multi-spin interactions, see also \cite {L+19} for recent developments.
  
 Fisher, see \cite{Fi65}, \cite{Fi67}, considered the case of complex $\beta$, again with the purpose of understanding the phase transition in the thermodynamic limit. 
 
 A straightforward change of variables transforms the partition function (1.4.1) into the expectation (1.1.2). Indeed, given $\phi_i: \{-1, 1\}^n \longrightarrow {\Bbb C}$ and $\beta$ as in (1.4.1), we define $\widehat{\phi}_i: \{0, 1\}^n \longrightarrow {\Bbb C}$ by 
 $$\widehat{\phi}_i \left(\xi_1, \ldots, \xi_n\right) = \beta \phi_i\left( 2\xi_1-1, \ldots, 2\xi_n-1\right).$$
  For the external field $\alpha$ and the inverse temperature $\beta$ in (1.4.1), we let 
 $$p={e^{-{\beta \alpha}}\over e^{\beta \alpha} + e^{-\beta \alpha}} \quad \text{and} \quad q={e^{\beta \alpha} \over e^{\beta \alpha} + e^{-\beta \alpha}} \tag1.4.3$$
 and consider the probability product measure in $\{0, 1\}^n$ defined by (1.1.1). Then 
 $$\sum\Sb x \in \{-1, 1\}^n: \\ x=\left(\xi_1, \ldots, \xi_n\right) \endSb \exp\left\{ \beta \sum_{i=1}^m \phi_i(x) - \beta \alpha \sum_{j=1}^n \xi_j \right\}
 =\left(e^{-\beta \alpha} + e^{\beta \alpha}\right)^n \EE \exp\left\{  \sum_{i=1}^m  \widehat{\phi}_i\right\}.$$
 We note that the expression 
 $$ \sum_{i:\ j \in J_i} L_i, \tag1.4.4$$
 used in Theorems 1.2 and 1.3, indicates the {\it influence} of the $j$-th spin $\xi_j$ onto the total energy of the system.
 
 Since we assume that $p \leq q$ in (1.1.1), we assume that $\alpha \geq 0$. Although we keep $\alpha$ real, we can formally handle complex $\alpha$ by introducing additional functions $\phi_j$ to absorb the imaginary parts of $\alpha \xi_j$.
  
  It has been long understood that the presence of a strong external field, that is, when $\alpha$ is large, makes the phase transition go away. As $\alpha$ grows, the Gibbs probability space collapses to the most likely configuration $\left(-1, \ldots, -1\right)$, minimizing the energy, while for the partition function we have
 $$Z(H, \beta) \sim \left(e^{\beta \alpha} + e^{-\beta \alpha}\right)^n. \tag1.4.5$$
 The task then is to give some quantitative bounds to ensure that  ``$\sim$'' in (1.4.5) also results in $Z(H, \beta) \ne 0$.
 
 The oldest and by far the most popular method is that of {\it cluster expansion}, see \cite{Br86}, Chapter 5 of \cite{FV18} and also \cite{Je24} for a recent survey
 from a combinatorial point of view. It is a perturbative technique based on the series expansion of $\ln Z(H, \beta)$ about a limit point at which the partition function collapses to some simple expression, such as the right hand side of (1.4.5). First introduced by Mayer \cite{Ma37}, it has been widely applied, and is still being actively adapted to new situations, see, for example, \cite{C+26}.
 
 Theorems 1.2 and 1.3 are proved by an inductive argument. A related, but different inductive argument was used before by the author in \cite{Ba26} for a model that allowed arbitrary spin values, but no external field. Although the external field can be formally incorporated into $\phi_i$, the method of \cite{Ba26} does not seem to lead to Theorems 1.2 and 1.3. 
 
 Compared to the results obtained via the cluster expansion, Theorem 1.2 appears to achieve an improvement in several directions. 
 
 First, it indicates a regime where, for a fixed $\beta$, an {\it additive} increase in the external field $\alpha$, and hence in view of (1.4.3), a multiplicative change in $p$, allows for a {\it multiplicative} increase of the influence (1.4.4) of a spin that still keeps the partition function zero-free. In other words, the strength of the external field required to keep the partition function zero-free is logarithmic in the energy of interactions. It should be noted however, that this type of behavior can be observed only for a sufficiently large $r$.
 
 Second, the cluster expansion approach requires the values of $|\phi_i|$ to be uniformly small, cf. Chapters 5 and 6 of \cite{FV16}, while we impose a less restrictive condition of slowly-varying $\phi_i$, controlling their Lipschitz constant in the Hamming metric. 
 
 Third, the cluster expansion method appears to require $|\phi_i|$ to be exponentially small in the number $r$ of the arguments, see
Section 6.5.4 of \cite{FV18}, while in Theorem 1.2 the dependence of the Lipschitz constant of $\phi_i$ on $r$ is $\sim r^{-1/2}$.
 
As the external field $\alpha$ grows, so by (1.4.3) the value of $p$ gets smaller, we find ourselves under the circumstances of Theorem 1.3. There a linear increase in the influence of a spin requires a linear increase in the external field, to keep the partition function zero-free. This result is similar to the results obtained via the cluster expansion method. However, we still keep the advantages of controlling the Lipschitz constant, as opposed to the actual values, of $\phi_i$, and a modest dependence on $r$.
 
\subhead (1.5) Computing the partition function \endsubhead Theorems 1.2 and 1.3 allow one to efficiently approximate the values of the expectation (1.1.2) and, consequently, of the partition function (1.4.1). We sketch how it is done here, providing more details in Section 2. We say that two complex numbers $z_1 \ne 0$ and $z_2 \ne 0$ approximate each other within relative error $\epsilon >0$ if we can write $z_1=e^{w_1}$ and $z_2=e^{w_2}$ for some 
$w_1, w_2 \in {\Bbb C}$ such that $|w_1 - w_2| \leq \epsilon$.

Let functions $\phi_i: \{0, 1\}^n \longrightarrow {\Bbb C}$, their Lipschitz constants $L_i$, probabilities $p$ and $q$ and a constant $L$ be as in Theorems 1.2 and 1.3. Without loss of generality, we assume that 
$$\phi_i(0, \ldots, 0)=0 \quad \text{for} \quad i=1, \ldots, m, \tag1.5.1$$ since adding a constant $c$ to some $\phi_i$ results in multiplying 
the expectation (1.1.2) by $e^{c}$. Let us fix some number $\delta >0$ and suppose that the Part (2) in Theorem 1.2, respectively Theorem 1.3 is satisfied with some slack $2\delta$, namely we have 
$$(1+2\delta) \sum_{i:\ j \in J_i} L_i \ \leq \ {1 \over 10 \sqrt{rp}} \quad \text{for} \quad j=1, \ldots, n \tag1.5.2$$
in Theorem 1.2 and 
$$(1+2\delta) \sum_{i:\ j \in J_i} L_i \ \leq \ {1 \over 5} L \quad \text{for} \quad j=1, \ldots, n \tag1.5.3$$
in Theorem 1.3. 

Let 
$$M=3\exp\left\{ (1+2 \delta) r \sum_{i=1}^m L_i \right\}. \tag1.5.4$$
It turns out that for any $0 < \epsilon < 1$, one can approximate (1.1.2) in polynomial time from the expectations 
$$\EE \left( \phi_{i_1} \cdots \phi_{i_l}\right) \quad \text{where} \quad l \ \leq \ k=O_{\delta} \left( \ln \ln M - \ln \epsilon \right), \tag1.5.5$$
where the implied constant in the ``$O$" notation depends on $\delta$ alone. 

Once we have an access to computing $\phi_i(x)$ for a given $i=1, \ldots, m$ and $x \in \{0, 1\}^n$, each of the not more than $m^{k+1}$ expectations (1.5.5) can be computed in a straightforward way in $O\left(2^{rk}\right)$ time (recall that each $\phi_i$ depends on at most $r$ coordinates). If $r$ is fixed in advance, in view of (1.5.4)--(1.5.5), we obtained an algorithm of a quasi-polynomial complexity.

\subhead (1.6) Combinatorial connections \endsubhead Many of the same partition functions considered in statistical physics are also considered in combinatorics and computer science, often under different names. For example, what is known as the hard core model in physics, is known as independent sets in graphs in combinatorics, cf. \cite{Je24}, and what is known as the  monomer-dimer partition function in physics \cite{HL72}, is known as the matching polynomial in combinatorics \cite{LP09}. 
 
For a combinatorially defined family $\FF \subset 2^{\{1, \ldots, n\}}$, we would like to compute efficiently some statistics on $\FF$. One can identify the hard and soft approaches.
 
 Within the {\it hard} approach, we seek to compute or approximate  the univariate polynomial
 $$h_{\FF}(\lambda)=\sum_{A \in \FF} \lambda^{|A|}. \tag1.6.1$$
 Clearly, $h_{\FF}(1)=|\FF|$. Computing $h_{\FF}(\lambda)$ for $\lambda > 1$ gives some information on the sets $A \in \FF$ of large cardinality, while computing $h_{\FF}(\lambda)$ for $0 < \lambda <1$ gives some information on the sets $A \in \FF$ of small cardinality.
  
 Within the {\it soft} approach, we introduce a penalty function $w: 2^{\{1, \ldots, n\}} \longrightarrow {\Bbb R}$ such that $w(A)=0$ if $A \in \FF$ and $w(A) >0$ if $A \notin \FF$, and seek to approximate 
 $$s_{\FF}(\lambda) = \sum_{A \subset \{1, \ldots, n\} } e^{-\lambda w(A)}. \tag1.6.2$$
 Clearly, 
 $$\lim_{\lambda \longrightarrow +\infty} s_{\FF}(\lambda)=|\FF|.$$
 Often, we are able to compute (1.6.2) only for moderate values of $\lambda >0$, which still gives some information regarding $\FF$ (for example, an upper bound for $|\FF|$). 
 
 As an example, we consider the monomer-dimer (matchings in graphs) problem and its possible modification to matchings in hypergraphs (polymers).
 
 Let $G=(V, E)$ be an undirected simple graph with set $V$ of vertices, set $E$ of edges, without loops or multiple edges. A collection $F \subset E$ of edges 
 is called a {\it matching} if every vertex of $G$ is incident to at most one edge from $F$ (in particular, $F=\emptyset$ is a matching). Let $\FF \subset 2^E$ be the set of all matchings. The polynomial (1.6.1) is called the {\it matching polynomial} of $G$. The Heilmann - Lieb Theorem \cite{HL72} states that the roots of 
 $h_{\FF}$ are negative real, and, moreover, if the degrees of the vertices of $G$ do not exceed some $\Delta > 1$, then the roots of $h_{\FF}$ do not exceed 
 $-{1 \over 4 (\Delta-1)}$, see also \cite{Go81}.  Based on this, for any $\Delta$, fixed in advance, one can construct a polynomial time algorithm approximating the number $h_{\FF}(1)$ of matchings in a given graph within any given relative error $0 < \epsilon <1$ \cite{PR17}, see also \cite{Ba19}. We also note that a Markov Chain Monte Carlo based {\it randomized} polynomial time approximation algorithm was constructed earlier by Jerrum and Sinclair \cite{JS89}, without any assumption about the maximum degree $\Delta$. 
 
 Let $H=(V, E)$ be a hypergraph with set $V$ of vertices and set $E$ of edges. We assume that each edge $S \in E$ of $H$ is a subset $S \subset V$ with $|S| \leq k$. A collection $x \subset E$ of edges is called a {\it matching} if every vertex of $G$ is incident to at most one edge from $x$ (in particular, $x=\emptyset$ is a matching). Let $\FF$ be the set of all matchings in $H$. An analogue of the matching polynomial $h_{\FF}$ for hypergraphs lacks the real-rootedness property of its graph version, see also Section 5.5 of 
 \cite{Ba16} for some information on its roots. Below we sketch a soft approach to constructing the partition function (1.6.2) for hypergraph matchings.
 
 We identify all possible collections $x$ of edges of $H$ with the cube $\{0, 1\}^{E}$. Hence the coordinates of $\{0, 1\}^E$ are indexed by the edges $S$ of $H$, the coordinate of $x_S$ is 1 if $S \in x$ and 0 if $S \notin x$. With a vertex $v \in V$, we associate a function $\phi_v: \{0, 1\}^E \longrightarrow {\Bbb R}$ by 
 $$\phi_v(x)=\cases 0 &\text{if $v$ is contained in at most 1 edge from $x$} \\ s-1 &\text{ if $v$ is contained in $s \geq 2$ edges from $x$.} \endcases$$
 We note that $\phi_v$ is $1$-Lipschitz in the Hamming metric of $\{0, 1\}^{E}$. We make $\{0, 1\}^{E}$ a probability space as in (1.1.1), so that 
 $0 < p \leq 1/2$ is the probability that an edge $S$ is selected in $x$ and $q=1-p$ is the probability that it is not selected. For $\lambda \in {\Bbb C}$, we define 
 $$f(\lambda) = \EE \exp\left\{ -\lambda \sum_{v \in V}  \phi_v \right\}.$$
 For real $\lambda > 0$, each subset $x \subset E$ accounts in $f(\lambda)$ with weight 1 if $x$ is a matching, and with a weight exponentially small in the number of vertices where the matching condition is violated, and also exponentially small in the number of edges violating the matching condition at each vertex. 
 The influence (1.4.4) 
 of each coordinate $x_S$ does not exceed $k|\lambda|$, since every edge $S$ of $H$ contains at most $k$ vertices.
 
 It follows from Theorems 1.2 and 1.3 that as the probability $p$ of selecting an edge gets smaller, the radius $\rho$ of the zero-free disc $|\lambda| \leq \rho$
 for $f(\lambda)$ gets larger. Consequently, as we discuss in Section 1.5, the value of $\lambda >0$ for which $f(\lambda)$ can be efficiently approximated,
 also increases. Selecting an edge with probability $p$ allows us to zoom in on the collections of about $p |E|$ edges, so as we focus on smaller collections, we can charge larger penalty for violating the matching condition.
 
\subhead (1.7) Notation \endsubhead We denote by $\ii$ the imaginary unit, so that $\ii^2=-1$. For a complex number $z=u +\ii v$, where $u, v \in {\Bbb R}$, by $\Re$ and $\Im$ we denote the real and, respectively, imaginary part of $z$: $\Re\thinspace z =u$ and $\Im\thinspace z=v$.

 \head 2. Approximating the partition function \endhead

It has been recognized for some time now, that the complex zeros of a partition function are closely related to the computational complexity of approximating its value, see \cite{Ba16}, \cite{PR17}, \cite{B+21}, \cite{G+22}. The case where the partition function is a polynomial in some natural parameter was treated in detail in \cite{Ba16}, \cite{Ba19} and \cite{PR17}, among other works. If the partition function is not a polynomial, it can be approximated by a polynomial sufficiently closely (this approach was used, for example, in \cite{Ba26}). Here we describe a more direct approach that can be applied to the expectation (1.1.2).

For $\rho >0$, let 
$${\Bbb D}_{\rho}=\bigl\{z \in {\Bbb C}: \quad |z| \ \leq \ \rho \bigr\}$$
be the closed disc in the complex plane, centered at $0$ and of radius $\rho >0$. We start with the following result.
\proclaim{(2.1) Lemma} 
Let $\delta >0$ and let  $g: {\Bbb D}_{1+2\delta} \longrightarrow {\Bbb C}$ be a holomorphic function such that $g(0)=1$ and 
$$0 \ < \ |g(z)| \ \leq \ M \quad \text{for all} \quad z \in {\Bbb D}_{1+2\delta}$$
and for some $M \geq 1$. Since $g(z) \ne 0$ for all $z \in {\Bbb D}_{1+2\delta}$, we can choose a continuous branch of $h(z)=\ln g(z)$, and we choose it so that 
$h(0)=\ln 1=0$. For $k \geq 1$, let 
$$T_k(h; z)=\sum_{l=1}^k {h^{(l)}(0) \over l!} z^l$$
be the Taylor polynomial of $h$ of degree $k$, computed at $0$. Then 
$$\left| h(1) - T_k(h, 1)\right| \ \leq \ {2 \ln M \over \delta^2 (1+\delta)^{k-1}}.$$
Moreover, given $g^{(l)}(0)$, $l=1, \ldots, k$, one can compute $h^{(l)}(0)$, $l=1, \ldots, k$, in $O(k^2)$ time.
\endproclaim
\demo{Proof} We have
$$\Re\thinspace h(z) \ \leq \ \ln M \quad \text{for all} \quad z \in {\Bbb D}_{1+ 2\delta}. \tag2.1.1$$
By the Borel - Carath\'eodory Theorem, see for example, Section XII.3 of \cite{La99}, for all $0 < r < \rho$, we have 
$$\max_{z \in {\Bbb D}_r} |h(z)| \ \leq \ {2r \over \rho-r} \max_{z \in {\Bbb D}_{\rho}} \Re\thinspace h(z) + {\rho+r \over \rho-r} |h(0)|.\tag2.1.2$$
Choosing $\rho=1+2\delta$ and $r=1+\delta$, from (2.1.1) and (2.1.2) we obtain
$$|h(z)| \ \leq \ {2 + 2 \delta \over \delta} \ln M \quad \text{for all} \quad z \in {\Bbb D}_{1+\delta}. \tag2.1.3$$
Using the Cauchy bound, 
$$\left| h^{l}(0) \over l! \right| \ \leq \ r^{-l} \max_{z \in {\Bbb D}_r} |h(z)|,$$
see for example, Section III.7 of \cite{La99}, from (2.1.3) we obtain
$$\split &\left| h(1)- T_k(h; 1 )\right| \leq \sum_{l=k+1}^{\infty} \left| {h^{(l)}(0) \over l!} \right| \ \leq \ {(2 +  2\delta) \ln M \over \delta} \sum_{l=k+1}^{\infty} 
\left(1+\delta \right)^{-l} \\
&\qquad = {(2+2 \delta) \ln M \over \delta} {1 \over \delta (1+\delta)^k} = {2 \ln M \over \delta^2 (1+\delta)^{k-1}}, \endsplit$$
as required.

Since $h'(z)=g'(z)/g(z)$, we have $g'(z)=h'(z) g(z)$ and 
$$g^{(l)}(0)=\sum_{i=0}^{l-1}  {l-1 \choose i} h^{(l-i)}(0) g^{(i)}(0) \quad \text{where} \quad g^{(0)}(0)=g(0)=1. \tag2.1.4$$
The system (2.1.4) for $l=1, \ldots, k$ is a system of linear equations in $k$ unknown $h^{(1)}(0), \ldots, h^{(k)}(0)$ with an triangular $k \times k$ matrix having 
$g(0)=1$ on the diagonal, which can be solved in $O(k^2)$ time see, for example, Section 2.2.2 of \cite{Ba16} for details.
{\hfill \hfill \hfill} \qed 
\enddemo

As follows from Lemma 2.1, to approximate $h(1)$ within an additive error of $0 < \epsilon < 1$ and hence to approximate $g(1)=e^{h(1)}$ within a relative error of $\epsilon$, one needs to compute the derivatives $g^{(1)}(0), \ldots, g^{(k)}(0)$ for some $k= O_{\delta}\left(\ln \ln M - \ln \epsilon\right)$, where the implied constant in the ``$O$" notation depends only on $\delta$. 

\subhead (2.2) Approximating the partition function \endsubhead 
Let functions $\phi_i: \{0, 1\}^n \longrightarrow {\Bbb C}$, their respective Lipschitz constants $L_i$ for $i=1, \ldots, m$,  probabilities $p$ and $q$, and 
a number $L$ be as in Theorems 1.2 and 1.3 and assume that the conditions of the theorems are satisfied with some fixed slack $2 \delta >0$, so that 
(1.5.2) and (1.5.3) hold. Our goal is to approximate the expectation (1.1.2).

As in Section 1.5, without loss of generality, we assume that (1.5.1) holds. We define a function 
$g: {\Bbb D}_{1+2\delta} \longrightarrow {\Bbb C}$ by 
$$g(z)=\left( \EE \exp\left\{ z \sum_{i=1}^m \phi_i \right\} \right) \quad \text{for} \quad z \in {\Bbb D}_{1+2 \delta}.$$
The quantity we wish to approximate is 
$$g(1)= \EE \exp\left\{ \sum_{i=1}^m \phi_i \right\}.$$
We have $g(0)=1$ and by Theorems 1.2 and 1.3, we have $g(z)\ne 0$ for all $z \in {\Bbb D}_{1+2\delta}$. In view of (1.5.1) we have 
$$|g(z)| \leq M,$$
where $M$ is defined by (1.5.4) (and we pad $M$ by an extra factor of $3$ to keep $\ln \ln M$ non-negative). We are now in the situation of Lemma 2.1. The lemma asserts then that we can approximate $g(1)$ within relative error 
$0 < \epsilon <1$ from the derivatives $g^{(1)}(0), \ldots, g^{(k)}(0)$ with 
$$k=O_{\delta} \left( \ln \ln M  -\ln \epsilon\right). \tag2.2.1$$
We have 
$$g^{(l)}(0)=\EE \left( \sum_{i=1}^m  \phi_i \right)^l =\sum_{1 \leq i_1, \ldots, i_l \leq m}  \EE \left(\phi_{i_1} \cdots \phi_{i_l}\right).$$
Hence approximating of (1.1.2) within relative error $0< \epsilon <1$ reduces to computing not more than $m^{k+1}$ expectations 
$\EE \left(\phi_{i_1} \cdots \phi_{i_k}\right)$ with $l \leq k$. As we argued in Section 1.5, if each $\phi_i$ depends on at most $r$ coordinates for some $r$ fixed in advance, in view of (2.2.1) we obtain a quasi-polynomial algorithm.

\head 3. Preparations \endhead

The proofs of Theorems 1.2 and 1.3 are quite similar. We give a detailed proof of Theorem 1.2 and then indicate the changes that need to be made for Theorem 1.3. The main goal of this section is to prove the following result.
\proclaim{(3.1) Theorem} Let $\{0, 1\}^n$ be the Boolean cube with the probability measure defined by 
$${\Bbb P}(x)=\prod_{j=1}^n p^{\xi_j} q^{1-\xi_j} \quad \text{for} \quad x \in \{0, 1\}^n, \quad x=\left(\xi_1, \ldots, \xi_n\right),$$
where $0 < p \leq q$ and $p+q=1$. 
Let $f: \{0, 1\}^n \longrightarrow {\Bbb C}$ be an $L$-Lipschitz function in the Hamming metric of $\{0, 1\}^n$.
\roster
\item Suppose that $pn \geq 12$ and that
$$L={1\over 5\sqrt{pn}} $$
Then 
$$\left| \EE e^f \right| \ \geq \ {1 \over 2} \EE \left| e^f \right|.$$
\item Suppose that 
$$p={e^{-6L} \over n}.$$
 Then 
$$\left| \EE e^f \right| \ \geq \ {1 \over 5} \EE \left| e^f \right|.$$
\endroster
\endproclaim

The proof of Theorem 3.1 is based on several standard estimates, certainly known in some form, cf. \cite{Mc98}. To prove Theorem 3.1, we give complete proofs of the results in the form we need.

We start with the case of $n=1$ and real-valued Lipschitz functions on a two-point space.
\proclaim{(3.2) Lemma} For $L \geq 0$, let $f: \{0, 1\} \longrightarrow {\Bbb R}$ be an $L$-Lipschitz function,
where 
$${\Bbb P}(0)=q, \quad {\Bbb P}(1)=p,$$
and $0 < p \leq q$, $p+q=1$.
Then 
$$\EE e^f \ \leq \ \left( p e^{Lq} + q e^{-Lp} \right)  \exp\left\{ \EE f \right\}.$$
\endproclaim
\demo{Proof} The statement is trivially true for $L=0$, so without loss of generality we assume that $L>0$.
Replacing $f$ by $f -\EE f$, without loss of generality we assume that $\EE f=0$. Then 
$$f(0)=-\lambda p \quad \text{and} \quad f(1)=\lambda q$$
for some $\lambda \in {\Bbb R}$, in which case the Lipschitz constant of $f$ is $|\lambda|$ and we have 
$$\EE e^f = p e^{\lambda q} + q e^{-\lambda p}.$$
Since $f$ is $L$-Lipschitz, we must have 
$$\EE e^f \ \leq \ \max_{\lambda:\ |\lambda| \leq L} \left(p e^{\lambda q} + q e^{-\lambda p}\right). \tag3.2.1$$
We define $h: {\Bbb R} \longrightarrow {\Bbb R}$ by $h(\lambda)=p e^{\lambda q} +q e^{-\lambda p}$.
Then 
$$h'(\lambda)=pq e^{\lambda q} - pq e^{-\lambda p} = pq\left(e^{\lambda q} - e^{-\lambda p}\right),$$
from which it follows that $h(\lambda)$ is increasing for $\lambda > 0$ and decreasing for $\lambda < 0$.
Therefore, from (3.2.1) we conclude that 
$$\EE e^f \ \leq \ \max\left\{h(L),\ h(-L) \right\} = \max\left\{p e^{Lq} + q e^{-Lp},\ pe^{-Lq} + qe^{Lp} \right\}. \tag3.2.2$$
We have 
$$\split h(L)-h(-L)=&\left(p e^{Lq} + q e^{-Lp}\right) - \left(pe^{-Lq} +q e^{Lp}\right)\\=&p\left(e^{Lq} - e^{-Lq}\right) -q \left(e^{Lp} - e^{-Lp}\right).\endsplit \tag3.2.3$$
Next, we consider a function $u: {\Bbb R}_+ \longrightarrow {\Bbb R}$ defined by 
$$u(t)=\ln t - \ln \left(e^t -e^{-t}\right) \quad \text{for} \quad t >0.$$
We have
$$u'(t)={1 \over t} - {e^t + e^{-t} \over e^t - e^{-t}}={1 \over t} - {1 \over \tanh t}.$$
Since 
$$\left( \tanh t\right)'= \left({ e^t - e^{-t} \over e^t + e^{-t}}\right)'={(e^t + e^{-t})^2 - (e^t - e^{-t})^2 \over (e^t +e^{-t})^2}=1- \left(\tanh t\right)^2 < 1$$
and $\tanh(0)=0$, we have $\tanh t < t$ for all $t >0$. 
Therefore, 
$${1 \over \tanh t} \ > \ {1 \over t} \quad \text{and} \quad u'(t) < 0 \quad \text{for all} \quad t > 0.$$ Hence $u(t)$ is decreasing for $t > 0$. Therefore, the function 
$$\ln t - \ln \left(e^{Lt} - e^{-Lt}\right)=u(Lt) - \ln L$$
is also decreasing for all $t >0$. Since $p \leq q$, we have 
$$\ln p - \ln \left( e^{Lp} - e^{-Lp} \right) \ \geq \ \ln q - \ln \left(e^{Lq} - e^{-Lq} \right)$$
and 
$$p \left(e^{Lq} - e^{-Lq}\right) \ \geq \ q \left(e^{Lp} - e^{-Lp}\right).$$
Hence by (3.2.3) we obtain that $h(L) \geq h(-L)$ and by (3.2.2) we complete the proof.
{\hfill \hfill \hfill} \qed 
\enddemo

\proclaim{(3.3) Lemma} Let $\{0, 1\}^n$ be the Boolean cube, endowed with the probability measure as in Theorem 3.1. Let 
$f: \{0, 1\}^n \longrightarrow {\Bbb R}$ be an $L$-Lipschitz function, such that $\EE f=0$. Then
$$\EE e^f \ \leq \ \left(p e^{Lq} + q e^{-Lp}\right)^n.$$
\endproclaim
\demo{Proof} The proof is by a standard martingale argument, cf. \cite{Mc98}. We proceed by induction on $n$. For $n=1$, the result is given by Lemma 3.2.
Suppose now that $n \geq 2$. We define a function $g: \{0, 1\}^{n-1} \longrightarrow {\Bbb R}$ by 
$$g\left(\xi_1, \ldots, \xi_{n-1}\right)= pf\left(\xi_1, \ldots, \xi_{n-1}, 1\right) + qf\left(\xi_1, \ldots, \xi_{n-1}, 0 \right).$$
Writing $x \in \{0, 1\}^n$ as $x=(y, \xi_n)$ for $y \in \{0, 1\}^{n-1}$ and $\xi_n = \pm 1$ and denoting by $\EE_x$ the expectation in $\{0, 1\}^n$ and by 
$\EE_y$ the expectation in $\{0, 1\}^{n-1}$, we obtain
$$\EE_y g = \EE_x f =0.$$
In addition, $g$ is $L$-Lipschitz and hence from the induction hypothesis, 
$$\EE_y e^g \ \leq \ \left(p e^{Lq} + q e^{-Lp}\right)^{n-1}. \tag3.3.1$$
On the other hand, applying Lemma 3.2, we get
$$\split \EE_x e^f = &\EE_y \left(pe^{f(y,1)} + q e^{f(y,0)}\right) \ \leq \ \EE_y \left(\left( p e^{Lq} + q e^{-Lp}\right) e^{g(y)}\right) \\ = 
&\left( p e^{Lq} + q e^{-Lp}\right) \EE_y e^g \ \leq \ \left(p e^{Lq} + q e^{-Lp}\right)^n, \endsplit$$
where the last inequality follows by (3.3.1).
{\hfill \hfill \hfill} \qed
\enddemo

\proclaim{(3.4) Corollary} Let $f: \{0, 1\}^n \longrightarrow {\Bbb R}$ be an $L$-Lipschitz function such that $\EE f=0$, as in Lemma 3.3.
\roster
\item If $L \leq 1/q$ then 
$$\EE e^f \ \leq \ \exp\left\{ {3 L^2 pq n \over 4} \right\}.$$
\item Suppose that $p =\alpha/n$ for some $0 < \alpha \leq 1$. Then 
$$\EE e^f \ \leq \ \exp\left\{ \alpha e^L \right\}.$$
\endroster
\endproclaim
\demo{Proof}  Using that 
$$e^x \ \leq \ 1 + x + {3 \over 4} x^2 \quad \text{and} \quad e^{-x} \ \leq \ 1 - x + {3 \over 4} x^2 \quad \text{for} \quad 0 \leq x \leq 1,$$
from Lemma 3.3 in Part (1) we get 
$$\split \EE e^f \ \leq \ &\left(p e^{Lq} + q e^{-Lp}\right)^n \ \leq \ \left(p \left(1+ Lq + {3 \over 4} L^2 q^2\right) + q \left(1-L p + {3 \over 4} L^2 p^2 \right)\right)^n \\
=&\left(1 + {3 \over 4} L^2 \left(pq^2 + q p^2\right)\right)^n = \left(1 + {3L^2 pq \over 4}  \right)^n \ \leq \ \exp\left\{ {3 L^2 pq n \over 4} \right\},\endsplit$$
and Part (1) follows.

In Part (2), from Lemma 3.3 we get 
$$ \EE e^f \ \leq \ \left(p e^{Lq} + q e^{-Lp}\right)^n \ \leq \ \left({\alpha \over n} e^L + 1\right)^n \ \leq \ \exp\left\{ \alpha e^L \right\}.$$
{\hfill \hfill \hfill} \qed
\enddemo

The following result is a standard application of the Laplace transform method.

\proclaim{(3.5) Lemma} Let $\{0, 1\}^n$ be the Boolean cube, endowed with the probability measure as in Theorem 3.1. Let 
$f: \{0, 1\}^n \longrightarrow {\Bbb R}$ be a $1$-Lipschitz function such that $\EE f=0$ and let $t > 0$ be a real number.
\roster
\item 
For $t \leq 3pn/2$, we have
$${\Bbb P}(f \geq t) \ \leq \ \exp\left\{-{t^2\over 3pqn}\right\}.$$
\item Suppose that $p=\alpha/n$ for some $0 < \alpha \leq 1$ and that $t > \alpha$. Then 
$${\Bbb P}(f \geq t) \ \leq \ \exp\left\{t \left(1+ \ln {\alpha \over t}\right) \right\}.$$
\endroster
\endproclaim
\demo{Proof} For $\lambda >0$ from the Markov inequality, we obtain
$${\Bbb P}(f \geq t) = {\Bbb P}\left(e^{\lambda f} \geq e^{\lambda t}\right) \ \leq \ e^{-\lambda t} \EE e^{\lambda f}. \tag3.5.1$$
To prove Part (1), we choose 
$$\lambda = {2 t \over 3pq n} \ \leq \ {1 \over q}.$$
Then $\lambda f$ is $\lambda$-Lipschitz, so from (3.5.1) and Part (1) of Corollary 3.4, we get 
$$\split {\Bbb P}(f \geq t) \ \leq \ &\exp\left\{-{2t^2 \over 3pqn} \right\} \exp\left\{ {3 \lambda^2 pq n \over 4}\right\} 
=\exp\left\{ -{2 t^2 \over 3pqn} +{t^2 \over 3pqn}\right\} \\= &\exp\left\{ -{t^2 \over 3pqn}\right\}, \endsplit$$
as required.

To prove Part (2), in (3.5.1) we choose $\lambda = \ln t - \ln \alpha >0$. Since $\lambda f$ is $\lambda$-Lipschitz, from (3.5.1) and Part (2) of Corollary 3.4, we obtain
$$\split {\Bbb P}(f \geq t) \ \leq \ &\exp\left\{ -t \ln t + t \ln \alpha \right\} \exp\left\{ \alpha \exp\left\{ \ln t - \ln \alpha\right\} \right\} \\
=&\exp\left\{ -t \ln t + t \ln \alpha +t\right\} = \exp\left\{ t \left(1 + \ln {\alpha \over t} \right)\right\}. \endsplit $$
{\hfill \hfill \hfill} \qed
\enddemo

We are now ready to prove Theorem 3.1.
\subhead (3.6) Proof of Theorem 3.1 \endsubhead Replacing $f$ by $f - \EE f$, without loss of generality, we assume that $\EE f=0$. Let 
$f=g+\ii h$, where $g, h: \{0, 1\}^n \longrightarrow {\Bbb R}$ are real-valued functions. Then $g$ and $h$ are $L$-Lipschitz and 
$\EE g=\EE h=0$. By the Jensen inequality,
$$\EE e^g \ \geq \ 1. \tag3.6.1$$ 
Let 
$$X=\left\{x \in \{0, 1\}^n: \ |h(x)| \ \leq \ 1 \right\} \quad \text{and} \quad \overline{X} =\{0, 1\}^n \setminus X.$$
We introduce two quantities,
$$a=\sum_{x \in X} e^{g(x)} {\Bbb P}(x) \ \geq \ 0 \quad \text{and} \quad b=\sum_{x \in \overline{X}} e^{g(x)} {\Bbb P}(x) \geq \ 0. \tag3.6.2$$
From (3.6.1), we have 
$$a+b \ \geq \ 1. \tag3.6.3$$
Clearly, we have 
$$\EE \left| e^f \right| = \EE e^g = a+b. \tag3.6.4$$
Also,
$$\aligned \left| \EE e^f \right| \ \geq \ &\left| \sum_{x \in X} e^{f(x)} {\Bbb P}(x) \right| - \left| \sum_{x \in \overline{X}} e^{f(x)} {\Bbb P}(x) \right| \\ \geq \ 
&\Re \left(\sum_{x \in X} e^{g(x)+\ii h(x)} {\Bbb P}(x) \right) -  \sum_{x \in \overline{X}} \left| e^{g(x)+\ii h(x)}\right| {\Bbb P}(x) \\
=&\sum_{x \in X} e^{g(x)} \cos h(x) {\Bbb P}(x) -  \sum_{x \in \overline{X}} e^{g(x)} \\
\ \geq \ & a (\cos 1) - b.\endaligned \tag3.6.5$$

If $b < 0.5$, from (3.6.3) we have $a \geq 1-b > 0.5$ and then from (3.6.4) and (3.6.5), we get 
$${\left| \EE e^f \right| \over \EE \left| e^f \right|} \ \geq \ {a (\cos 1) - b \over a+b} ={(\cos 1) - b/a \over 1+ b/a} \ \geq \ {(\cos 1) - b/(1-b) \over 1+ b/(1-b)} =
(1-b)( \cos 1) - b,$$
so in the end
$${\left| \EE e^f \right| \over \EE \left| e^f \right|} \ \geq \ (1-b)( \cos 1) - b \quad \text{provided} \quad b < 0.5. \tag3.6.6$$
To bound $b$, let $[\overline{X}]: \{0, 1\}^n \longrightarrow \{0, 1\}$ be the indicator of $\overline{X}$, 
$$[\overline{X}](x)=\cases 1 &\text{if\ } x \in \overline{X} \\ 0 &\text{if\ } x \in X. \endcases$$
Then, from (3.6.2) and the Cauchy - Schwarz inequality,
$$b=\EE \left( [\overline{X}] e^g\right) \ \leq \ \left(\EE [\overline{X}]^2 \right)^{1/2} \left(\EE e^{2g}\right)^{1/2} =
\left({\Bbb P}(\overline{X})\right)^{1/2} \left(\EE e^{2g}\right)^{1/2}. \tag3.6.7$$
We have 
$$\split {\Bbb P}(\overline{X}) = &{\Bbb P}(h > 1) + {\Bbb P}(h < -1) = {\Bbb P} \left(L^{-1} h > L^{-1}\right) + {\Bbb P}\left(L^{-1} h < -L^{-1}\right)\\=
&{\Bbb P} \left(L^{-1} h > L^{-1}\right) + {\Bbb P}\left(-L^{-1} h > L^{-1}\right). \endsplit$$
We note that $L^{-1} h$ and $-L^{-1} h$ are $1$-Lipschitz functions.

In Part (1), we have 
$$L^{-1} = 5 \sqrt{pn} \ \leq \ {3pn \over 2}, $$
since $pn \geq 12$. Applying Part (1) of Lemma 3.5, we conclude that 
$${\Bbb P}(\overline{X}) \ \leq \ 2 \exp\left\{ -{25 \over 3q} \right\} \ < \ 2 e^{-25/3}. \tag3.6.8$$
Since the function $2g$ is ${2 \over 5\sqrt{pn}}$-Lipschitz, by Part (1) of Corollary 3.4, we have 
$$\EE e^{2g} \ \leq \ \exp\left\{ {3 q \over 25 }\right\} \ < \ e^{3/25}. \tag3.6.9$$
Therefore, from (3.6.7)--(3.6.9), we have 
$$b \ \leq \ \sqrt{2} \exp\left\{ -{25 \over 6} + {3 \over 50}\right\} = \sqrt{2} \exp\left\{ -{308 \over 75}\right\} \approx 0.02328157330.$$
and from (3.6.6), we obtain 
$${\left| \EE e^f \right| \over \EE \left| e^f \right|} \ \geq \ \left(1- \sqrt{2} e^{-308/75}\right) (\cos 1) - \sqrt{2} e^{-308/75} \approx 0.5044416449 > 0.5,$$ 
as required.

In Part (2), noting that $L^{-1} > e^{-6L}$ for $L>0$ and applying Part (2) of Lemma 3.5,
we get 
$${\Bbb P}(\overline{X}) \ \leq \ 2\exp\left\{{1 + \ln \left(L e^{-6L}\right) \over L} \right\}.$$
From Part (2) of Corollary 3.4, we have 
$$\EE e^{2g} \ \leq \ \exp\left\{ e^{-6L} e^{2L} \right\} = \exp\left\{ e^{-4L}\right\}.$$
Therefore, form (3.6.7)--(3.6.9), we have 
$$b \ \leq \ \sqrt{2} \exp\left\{  {1 + \ln \left(L e^{-6L}\right) \over 2L}  + {e^{-4L} \over 2}\right\} =
\sqrt{2} \exp\left\{-3 + {1 + \ln L \over 2L} + {e^{-4L} \over 2} \right\}.$$
The maximum value of the function 
$$L \longmapsto {1 + \ln L \over 2L} \quad \text{for} \quad L \geq 0$$
is attained at the unique critical point $L=1$ and equal to $1/2$. Therefore,
$$b \ \leq \ \sqrt{2} e^{-2} \approx 0.1913929929$$
and from (3.6.6), we obtain 
$${\left| \EE e^f \right| \over \EE \left| e^f \right|} \ \geq \ \left(1- \sqrt{2} e^{-2}\right) (\cos 1) - \sqrt{2} e^{-2} \approx 0.2454992376 \ > \ {1 \over 5}. $$
{\hfill \hfill \hfill} \qed

\head 4. Proof of Theorem 1.2 \endhead 

We proceed by induction on $m$ and prove the following

\subhead (4.1) Claim \endsubhead Let $\phi_i: \{0, 1\}^n \longrightarrow {\Bbb C}$ be functions with respective Lipschitz constants $L_i$, $i=1, \ldots, m$, as in Theorem 1.2, and let $\widehat{\phi}_m: \{0, 1\}^n \longrightarrow {\Bbb C}$ be yet another $L_m$-Lipschitz function that depends on a subset of or all the coordinates that $\phi_m$ depends on, and 
such that 
$$\left| \widehat{\phi}_m(x) - \phi_m(x) \right| \ \leq \ \tau \quad \text{for some} \quad \tau \geq 0  \quad \text{and all} \quad x \in \{0, 1\}^n. \tag4.1.1$$
Then 
$$\EE \left\{ \phi_m + \sum_{i=1}^{m-1} \phi_i \right\} \ne 0, \quad \EE \left\{ \widehat{\phi}_m + \sum_{i=1}^{m-1} \phi_i \right\} \ne 0 \tag4.1.2$$
and the ratio of the two numbers (4.1.2) can be written as $e^{\alpha}$ for some $\alpha \in {\Bbb C}$ such that $|\alpha| \leq 2\tau$.

\subhead (4.2) Base $m=1$ \endsubhead We write $\phi$, $\widehat{\phi}$ and $L$ respectively, instead of $\phi_1$, $\widehat{\phi}_1$ and $L_1$. Let $J=J_1$, so that the function $\phi$ depends on the coordinates $\left\{ \xi_j:\ j \in J \right\}$, the function $\widehat{\phi}$ depends on all or some of those coordinates, and $|J| \leq r$. We consider the Boolean cube $\{0, 1\}^J$ with the coordinates $\xi_j$ indexed by $j \in J$ and the probability product measure (1.1.1). Furthermore, we formally consider $\phi$ and $\widehat{\phi}$ as functions 
$\phi, \widehat{\phi}: \{0, 1\}^J \longrightarrow {\Bbb C}$.

For $0 \leq s \leq 1$, we define $\phi_s: \{0, 1\}^J \longrightarrow {\Bbb C}$ by 
$$\phi_s(x)=(1-s) \phi + s \widehat{\phi},$$
so that $\phi_0=\phi$, $\phi_1=\widehat{\phi}$ and $\phi_s$ is $L$-Lipschitz for all $0 \leq s \leq 1$. Since $\phi_s$ depends on at most $r$ coordinates, and 
$$L\ \leq \ {1 \over 10 \sqrt{rp}} \ < \ {1 \over 5 \sqrt{rp}} $$
by Part (1) of Theorem 3.1, we have 
$$\left| \EE \exp\left\{ \phi_s\right\} \right| \ \geq \ {1 \over 2} \EE \left| \exp\left\{ \phi_s \right\} \right|. \tag4.2.1$$
Since $\EE \exp\left\{ \phi_s \right\} \ne 0$, we can choose a continuous branch of the function
$$s \longmapsto \ln \EE \exp\left\{\phi_s \right\}$$ and write 
$$\aligned &\ln \EE \exp\left\{ \widehat{\phi} \right\} - \ln \EE \exp\left\{ \phi \right\} = \int_0^1 {d \over d s} 
\ln  \EE \exp\left\{ \phi_s \right\} \ d s \\ 
&\qquad = \int_0^1 {\EE (\widehat{\phi} - \phi) \exp\left\{  \phi_s \right\} \over \EE \exp\left\{ \phi_s \right\} } \ d s. \endaligned
\tag4.2.2$$
From (4.1.1) we have 
$$\left| \EE  (\widehat{\phi}-\phi) \exp\left\{ \lambda \phi_s \right\} \right| \ \leq \ \tau \EE \left| \exp\left\{ \phi_s \right\} \right|$$
and hence from (4.2.1) and (4.2.2), we conclude that
$$\left| \ln \EE \exp \{  \widehat{\phi} \} - \ln \EE \exp \{\phi \} \right| \ \leq \ 2 \tau,$$
which concludes the proof of the base case.

\subhead (4.3) Induction step $m-1 \Longrightarrow m$, $m \geq 2$ \endsubhead Let $J=J_m$, so that $\phi_m$ and $\widehat{\phi}_m$ depend on the coordinates $\left\{ \xi_j:\ j \in J\right\}$ and $|J| \leq r$. If $J =\emptyset$ then $\phi_m$ and $\widehat{\phi}_m$ are constants and hence 
$$\aligned &\EE \left\{  \phi_m + \sum_{i=1}^{m-1} \phi_i \right\} = \exp\left\{ \EE  \phi_m\right\} \left( \EE \exp\left\{ \sum_{i=1}^{m-1} \phi_i \right\} \right) \quad \text{and} \\
&\EE \left\{ \widehat{\phi}_m + \sum_{i=1}^{m-1} \phi_i \right\} = \exp\left\{ \EE  \widehat{\phi}_m\right\}  \left(\EE \exp\left\{ \sum_{i=1}^{m-1}  \phi_i \right\} \right). \endaligned $$
By the induction hypothesis, the expectations are non-zero and by (4.1.1) their ratio can be written as $e^{\alpha}$, where $|\alpha| \leq \tau$. 

Let $\overline{J}=\{1, \ldots, n\} \setminus J$. Suppose now that $\overline{J} = \emptyset$. Then $n \leq r$. Similarly to Section 4.2, for $0 \leq s \leq 1$, we introduce a function 
$f_s: \{0, 1\}^n \longrightarrow {\Bbb C}$ by 
$$f_s = (1-s) \phi_m + s \widehat{\phi}_m + \sum_{i=1}^{m-1} \phi_i.$$
Let $x', x'' \in \{0, 1\}^n$, $x'=\left(\xi_j'\right)$, $x''=\left(\xi_j''\right)$ be two points that differ in the $k$-th coordinate only.
Then 
$$\left|f_s(x') - f_s(x'')\right| \ \leq \ L_m + \sum\Sb 1 \leq i \leq m-1: \\ k \in J_i \endSb L_i = \sum_{i:\ k \in J_i}  L_i \ \leq \ {1 \over 10 \sqrt{rp}} \ \leq \ {1 \over 5 \sqrt{np}}.$$
Hence $f_s$ is ${1 \over 5 \sqrt{np}}$-Lipschitz and by Part (1) of Theorem 3.1 we have 
$$\left| \EE e^{f_s} \right| \ \geq \ {1 \over 2} \EE \left| e^{f_s} \right| > 0.$$
Arguing as in Section 4.2, we define a continuous branch of $s \longmapsto \ln \EE e^{f_s}$ and bound 
$$\split &\left| \ln \EE e^{f_1} - \ln \EE e^{f_0}\right| = \left| \int_0^1 \left( {d \over ds} \ln \EE e^{f_s}\right) \ ds \right| = \left| \int_0^1 {\EE \left(\left(\widehat{\phi}_m - \phi_m\right) e^{f_s} \right)\over \EE e^{f_s}} \ ds \right| \\ &\qquad \leq \ \int_0^1 {\EE\left( |\widehat{\phi}_m - \phi_m|\left| e^{f_s} \right|\right) \over \left| \EE e^{f_s} \right|} \ ds \ \leq \ 2\tau,\endsplit$$
which concludes the induction step in the case of $\overline{J}=\emptyset$.

Hence without loss of generality, we assume that $J \subset \{1, \ldots, n\}$ is a proper subset.
We consider the Boolean cubes $\{0, 1\}^J$ and $\{0, 1\}^{\overline{J}}$ with the coordinates $\xi_j$ indexed by $j \in J$ and $j \in \overline{J}$ respectively, and the probability product measures defined by (1.1.1). Let $I \subset \{1, \ldots, m-1\}$ be the set of all indices $1 \leq i \leq m-1$ such that $\phi_i$ depends on some of the coordinates $\xi_j$ with $j \in J$. For $i \in I$ and a point $x \in \{0, 1\}^J$, we define a function 
$\phi_i(\cdot\ | x): \{0, 1\}^{\overline J} \longrightarrow {\Bbb C}$ obtained by restricting $\phi_i$ to the points with the coordinates $\xi_j$ with $j \in J$ matching those of $x$. Note that the functions $\phi_i$ with $i \in \{1, \ldots, m-1\} \setminus I$ depend only on the coordinates $\xi_j$ with $j \in \overline{J}$, so we may consider them as functions on $\{0, 1\}^{\overline{J}}$.

We define $\Psi: \{0, 1\}^{J} \longrightarrow {\Bbb C}$ by 
$$\Psi(x)=\EE_{\overline{J}} \exp\left\{ \sum_{i \in I} \phi_i(\cdot\ | x) + \sum\Sb 1 \leq i \leq m-1 \\ i \notin I \endSb \phi_i \right\}, \tag4.3.1$$
where the expectation is taken with respect to the product measure (1.1.1) in $\{0, 1\}^{\overline{J}}$. Since each function $\phi_i(\cdot\ |x)$ is also $L_i$-Lipschitz, by the induction hypothesis, we have 
$$\Psi(x) \ne 0 \quad \text{for all} \quad x \in \{0, 1\}^{J}.$$
Furthermore, suppose that $x', x'' \in \{0, 1\}^J$ are two points that differ in one coordinate $\xi_k$ with $k \in J$ and let $I_k \subset I$ be the set of indices $i \in I$ such that $\phi_i$ depends on $\xi_k$.
Since $\phi_i$ are $L_i$-Lipschitz, 
we have 
$$ | \phi_i(\cdot\ | x') - \phi_i(\cdot\ | x'')| \ \leq \ L_i \quad \text{for all} \quad i \in I_k$$
and applying the induction hypothesis $|I_k|$ times, we conclude that 
$${\Psi(x') \over \Psi(x'')} = e^{\alpha} \quad \text{where} \quad |\alpha| \ \leq \ 2\sum_{i \in I_k} L_i = 2 \sum\Sb 1 \leq i \leq m-1: \\ k \in J_i \endSb L_i.$$
Next, we show that we can define $\psi(x)=\ln \Psi(x)$ in a consistent way, so that the bound for the ratio $\Psi(x')/\Psi(x'')$ translates into the bound for the difference 
of $\psi(x')-\psi(x'')$. The difficulty here is that $\ln$ is defined up to an additive term of $2\pi u \ii$ for $u \in {\Bbb Z}$.
\medskip
{\bf (4.3.2) Immediate goal:} We intend to construct a function $\psi: \{0, 1\}^{J} \longrightarrow {\Bbb C}$ such that $\Psi(x)=e^{\psi(x)}$ for all $x \in \{0, 1\}^J$ and 
$$ |\psi(x') - \psi(x'')| \ \leq \ 2\sum_{i \in I_k} L_i= 2 \sum\Sb 1 \leq i \leq m-1: \\ k \in J_i \endSb L_i,$$
provided $x', x'' \in \{0, 1\}^J$ differ in a single coordinate $\xi_k$ with $k \in J$. 
\medskip
To accomplish (4.3.2) we extend the map $\Psi$ defined by (4.3.1) to the solid cube $[0, 1]^J$. For $i \in I$ and $y \in [0, 1]^J$, we define 
$\phi_i(\cdot\ | y): \{0, 1\}^{\overline{J}} \longrightarrow {\Bbb C}$ by 
$$\split &\phi_i(\cdot\ | y) = \sum\Sb x \in \{0, 1\}^J: \\ x=\left(\xi_j, j \in J \right) \endSb \phi_i(\cdot\ | x) \prod_{j \in J} \eta_j^{\xi_j} \left(1-\eta_j\right)^{1-\xi_j},\\
&\qquad \text{where} \quad  y=\left(\eta_j:\ j \in J \right): \quad 0 \leq \eta_j \leq 1 \quad \text{for} \quad j \in J, \endsplit \tag4.3.3$$
and where we agree that $0^0=1$.
Denoting the extension of $\Psi$ also by $\Psi$, we define $\Psi: [0, 1]^J \longrightarrow {\Bbb C}$ by 
$$\Psi(y) = \EE_{\overline{J}} \exp\left\{ \sum_{i \in I} \phi_i (\cdot \ | y) + \sum\Sb 1 \leq i \leq m-1: \\ i \notin I \endSb \phi_i \right\} \quad \text{for} \quad y \in [0, 1]^J. \tag4.3.4$$
If $y \in \{0, 1\}^J$ then the definition (4.3.3) coincides with the previous definition of functions $\phi_i(\cdot\ | x)$ for $x=y \in \{0, 1\}^J$ and hence (4.3.4) is indeed an extension of (4.3.1). Moreover, for any $y \in [0, 1]^J$, the functions $\phi_i(\cdot\ | y)$ are $L_i$-Lipschitz, and hence by the induction hypothesis we have 
$\Psi(y) \ne 0$ for all $y \in [0, 1]^J$. Thus we can choose a continuous branch of $\psi(y) = \ln \Psi(y)$ for $y \in [0, 1]^J$. We observe that in each coordinate $\eta_j$ the function $\Psi(y)$ is affine, that is, when all other coordinates are fixed, $\Psi(y)$ is of the form $\alpha \eta_j + \beta$ for some $\alpha, \beta \in {\Bbb C}$. Hence as $\eta_j$ changes from $0$ to $1$, while all other coordinates remain fixed, the argument of $\Psi(y)$ continuously changes by a total angle less than $\pi$. It follows then 
that if $x', x'' \in \{0, 1\}^J$ differ in one coordinate only, we have 
$$\left|\Im\thinspace \psi(x') - \Im\thinspace \psi(x'') \right| \ < \ \pi$$
and the function $\psi$ satisfies the requirement of (4.3.2).

Since $\phi_m$ and $\widehat{\phi}_m$ depend on the coordinates $\xi_j$ with $j \in J$, we formally consider them as functions $\phi_m, \widehat{\phi}_m: \{0, 1\}^J \longrightarrow {\Bbb C}$. Then we can write
$$\aligned &\EE \exp\left\{  \phi_m + \sum_{i=1}^{m-1}  \phi_i \right\} = \EE_J \exp\left\{ \phi_m + \psi \right\} \quad \text{and} \\
&\EE \exp\left\{ \widehat{\phi}_m + \sum_{i=1}^{m-1} \phi_i \right\} = \EE_J \exp\left\{ \widehat{\phi}_m + \psi \right\},
\endaligned \tag4.3.5$$
where the expectation in the right hand side is taken with respect to the product measure (1.1.1) in $\{0, 1\}^J$. We conclude the proof in a similar way as in Section 4.2.
For $0 \leq s \leq 1$, we define $\tilde{\phi}_s: \{0, 1\}^J \longrightarrow {\Bbb C}$ by 
$$\tilde{\phi}_s=(1-s) \phi_m+ s \widehat{\phi}_m.$$
Then $\tilde{\phi}_s$ is $L_m$-Lipschitz and from (4.3.2), we conclude that $ \tilde{\phi}_s + \psi$ is $L$-Lipschitz with 
$$L =L_m + 2 \max_{j \in J}  \sum\Sb 1 \leq i \leq m-1 \\ j \in J_i \endSb L_i\ \leq \ 2 \max_{j \in J} \sum_{i:\ j \in J_i} L_i \ \leq \ {1 \over 5 \sqrt{rp}}.$$ Therefore, by Part (1) of Theorem 3.1, we have 
$$\left| \EE_J \exp\left\{\tilde{\phi}_{s} + \psi \right\} \right| \ \geq \ {1 \over 2} \EE_J \left| \exp \left\{ \tilde{\phi}_s + \psi \right\} \right|. \tag4.3.6$$
Since $\EE_J \exp\left\{  \tilde{\phi}_s + \psi \right\} \ne 0$ for $0 \leq s \leq 1$, we can choose a continuous branch of the function 
$s \longmapsto \ln \EE_J  \exp\left\{ \tilde{\phi}_s + \psi \right\}$.
Then we have 
$$\split &\ln \EE_J \exp\left\{ \widehat{\phi}_m + \psi \right\} - \ln \EE_J  \exp\left\{  \phi_m + \psi \right\} \\&\qquad = 
\int_0^1 {d \over ds} \ln \EE_J  \exp\left\{  \tilde{\phi}_s + \psi \right\}\ d \tau \\
&\qquad = \int_0^1 {\EE_J  (\widehat{\phi}_m - \phi_m) \exp \left\{  \tilde{\phi}_s + \psi \right\} \over \EE_J  \exp\left\{ \tilde{\phi}_s + \psi \right\}} \ d s. \endsplit \tag4.3.7$$
As in Section 4.2, using (4.1.1), we bound 
$$\left|  \EE_J  (\widehat{\phi}_m - \phi_m) \exp \left\{ \lambda_m \tilde{\phi}_s + \psi \right\}\right| \ \leq \ \tau \EE_J \left| \exp \left\{ \tilde{\phi}_s + \psi \right\}\right|.$$
It then follows from (4.3.6) and (4.3.7) that 
$$\left| \ln \EE_J \exp\left\{  \widehat{\phi}_m + \psi \right\} - \ln \EE_J  \exp\left\{ \phi_m + \psi \right\} \right| \ \leq \ 2 \tau,$$
and the proof follows by (4.3.5).
{\hfill \hfill \hfill} \qed 

\head 5. Proof of Theorem 1.3 \endhead

The proof of Theorem 1.3 is very similar to that of Theorem 1.2. We proceed by induction on $m$ and prove the following
\subhead (5.1) Claim\endsubhead For $i=1, \ldots, m$, let $\phi_i: \{0, 1\}^n \longrightarrow {\Bbb C}$ be functions with respective Lipschitz constants $L_i$ and let $L >0$ and $p$ be as in Theorem 1.3. Suppose that $\widehat{\phi}_m: \{0, 1\}^n \longrightarrow {\Bbb C}$ is yet another 1-Lipschitz function that depends on a subset or all of the coordinates that $\phi_m$ depends on and such that 
$$\left|\widehat{\phi}_m(x)-\phi_m(x) \right| \ \leq \ \tau \quad \text{for some} \quad \tau > 0 \quad \text{for all} \quad x \in \{0, 1\}^n.$$
Then 
$$\EE \exp\left\{ \phi_m + \sum_{i=1}^{m-1}  \phi_i \right\} \ne 0, \quad \EE \exp\left\{  \widehat{\phi}_m + \sum_{i=1}^{m-1} \phi_i \right\} \ne 0 \tag5.1.1$$
and the ratio of the numbers (5.1.1) can be written as $e^{\alpha}$, where $|\alpha| \leq 5 \tau$.

The proof proceeds exactly as in Section 4, except we use Part (2) of Theorem 3.1, instead of Part (1).

{\hfill \hfill \hfill} \qed

\enddocument
\end